\documentclass[11pt]{article}
\usepackage[a4paper,hmarginratio=1:1,
totalwidth=15.5cm,totalheight=22.7cm]{geometry}
\usepackage{bm,epstopdf,epsfig,amsmath,amssymb,amsfonts,colordvi,latexsym,comment,cancel,
verbatim}
\usepackage{graphicx,graphics}
\usepackage[font=md,captionskip=8pt]{subfig}

\newcommand{\eVdist}{\kern-0.06em}

\newcommand{\s}{\:\text{s}}

\newcommand{\be}{\begin{equation}}
\newcommand{\ee}{\end{equation}}
\newcommand{\bea}{\begin{eqnarray}}
\newcommand{\eea}{\end{eqnarray}}

\newcommand{\baa}{\begin{array}}
\newcommand{\eaa}{\end{array}}
\def\I{\mathrm{i}}
\long\def\symbolfootnote[#1]#2{\begingroup
\def\thefootnote{\fnsymbol{footnote}}\footnote[#1]{#2}\endgroup}
\begin{document}

\thispagestyle{empty}
\begin{flushright}
CERN-PH-TH/2012-175\\
\today
\end{flushright}

\vspace{2cm} 

\begin{center}
{\Large\bf Discrete Symmetries and Neutrino Mass Perturbations for $\theta_{13}$}
\vspace{1cm}

{\bf L. J. Hall$^{a}$\footnote{E-mail ljhall@lbl.gov} and G. G. Ross$^{b}$\footnote{E-mail g.ross1@physics.ox.ac.uk}
}

\bigskip

{\small $^a$ Berkeley Center for Theoretical Physics, Department of Physics }

{\small and Theoretical Physics Group, Lawrence Berkeley National Laboratory,}

{\small University of California, Berkeley, CA 94720, USA.}

{\small $^b$ Rudolf\, Peierls \,Centre \,for Theoretical Physics, University of Oxford,}

{\small 1 Keble Road, Oxford OX1 3NP, United Kingdom.}
\end{center}

\bigskip
\begin{abstract}\noindent
The recent measurement of the third lepton mixing angle, $\theta_{13}$, has shown that, although small compared to $\theta_{12}$ and $\theta_{23}$, it is much larger than anticipated in schemes that generate Tri-Bi-Maximal (TBM) or Golden Ratio (GR) mixing. We develop a model-independent formalism for perturbations away from exact TBM or GR mixing in the neutrino sector. Each resulting perturbation scheme reflects an underlying symmetry structure and involves a single complex parameter.   We show that such perturbations can readily fit the observed value of $\theta_{13}$, which is then correlated with a change in the other mixing angles. We also determine the implication for the lepton CP violating phases. For comparison we determine the predictions for Bi-Maximal mixing corrected by charged lepton mixing and we discuss the accuracy that will be needed to distinguish between the various schemes.
\end{abstract}

\newpage

\section{Introduction}\label{introduction}

The observation of a non-zero value for the lepton mixing angle, $\theta_{13}$, challenges models of charged lepton and neutrino masses based on Tri-Bi-Maximal (TBM) or Golden Ratio (GR) mixing.  To date five experiments, T2K\cite{Abe:2011sj}, MINOS\cite{Adamson:2011qu}, DOUBLE CHOOZ\cite{Abe:2011fz}, Daya Bay\cite{An:2012eh} and RENO\cite{Ahn:2012nd} have reported measurements of the mixing angle. Taken together with the other neutrino oscillation experimental measurements two groups have recently produced global fits to the PMNS matrix mixing angles and Dirac CP violating phase, the most recent one is presented in Table \ref{fits}. 

\begin{table}[h]
\caption{\label{Synopsis} Results for the lepton mixing angles and Dirac CP violating phase taken from the  global fit to neutrino oscillation data \cite{GonzalezGarcia:2012sz} (for previous fits see \cite{Fogli:2012ua,Tortola:2012te}). Separate results are shown for $\theta_{23}$ for the cases of Normal Hierarchy (NH) and Inverted Hierarchy (IH).}
\centering
\begin{tabular}{|c|ccc|}
\hline 
Parameter & Best fit & $1\sigma$ range  & $3\sigma$ range\\
\hline
$\sin^2 \theta_{12}/10^{-1}$  & 3.0 & 2.87 -- 3.13  & 2.7 -- 3.4 \\
\hline
$\sin^2 \theta_{13}/10^{-2}$  & 2.3 & 2.07 -- 2.53  & 1.6 -- 3.0 \\
\hline
$\sin^2 \theta_{23}/10^{-1}$ (NH) & 4.1 & 4.08 -- 4.14 & 3.4 -- 6.7 \\
$\sin^2 \theta_{23}/10^{-1}$  (IH) & 5.9 & 5.68 -- 6.11  & 3.4 -- 6.6 \\
\hline
$\delta/\pi$  & 1.7& 0.90 -- 2.03  & 0 -- 2 \\
\hline
\end{tabular}
\label{fits}
\end{table}

One may see that, at the $3\sigma$ level, $\sin^{2}\theta_{23}$ is in reasonable agreement with the TBM, GR and Bi-Maximal (BM) value of $1/2$. The value of $\sin^{2}\theta_{12}$ is also in reasonable agreement at the $3\sigma$ level with the TBM value of $1/3$ or the two proposed GR values of  $0.276$  and $0.345$ (but not with the BM value of 0.5) but the value of $\sin^{2}\theta_{13}$ is not consistent with the TBM, GR or BM value of $0$. 

However such models, based on a family symmetry, usually apply only to the contribution to the PMNS matrix from the neutrino mass matrix and there is an additional contribution from the mixing angle needed to diagonalise  the charged lepton mass matrix. If the latter is given by an underlying $SU(5)$ GUT, with the Georgi Jarlskog form \cite{Georgi:1979df} to accommodate the muon mass, one gets a contribution to $\sin^{2}\theta_{13}$ given by\footnote{This is the GUT-related analogue of the phenomenologically successful prediction for the Cabibbo angle, $\sin\theta_{C}=|\sqrt{m_{d}/{m_{s}}}-e^{i\delta}\sqrt{m_{u}/m_{c}}|$ that results from a texture zero in the (1,1) entry and symmetry in the 12/21 entries of the up and down quark mass matrix \cite{Gatto:1968zz}.} $\sin^{2}\theta_{13}=\frac{m_{e}}{m_{\mu}}\sim 0.5.10^{-2}$,  still inconsistent with the recent measurements at the $3\sigma$ level. 
Thus, in order to fit $\theta_{13}$, the model must be modified to obtain a larger contribution from the charged lepton sector or the neutrino sector or both. Modifications of the charged lepton sector consistent with an underlying Grand Unified Theory (GUT) have been suggested \cite{Antusch:2011qg,Antusch:2012fb,Marzocca:2011dh} by changing the Clebsch Gordon factors associated with the GUT from those assumed by Georgi and Jarlskog. However a disadvantage of these schemes is that they lose the phenomenologically successful prediction for the Cabibbo angle mentioned above, the equally successful prediction $|V_{td}/V_{ts}|=\sqrt{m_{d}/m_{s}}$ \cite{Hall:1993ni} as well as the prediction for the charged leptons $Det \, M_{d}=Det \, M_{l} $ that also follows from texture zeros.  Minimal forms for the modifications from the charged lepton sector have also been studied \cite{Marzocca:2013cr}.

The alternative is to modify the neutrino sector. In this context one may note that TBM and GR models lead to a discrepancy in the small mixing angle, while providing a reasonable approximation to the large mixing angles. For this reason it is reasonable to consider small perturbations as the dominant origin of $\theta_{13}$\footnote{For recent reviews see \cite{King:2013eh,Altarelli:2010gt} and extensive references therein.}.  Our approach is to allow for mixing between the unperturbed neutrino mass eigenstates, concentrating on the bilinear mixing patterns that can keep small the deviation of the good TBM or GR predictions for  $\theta_{12}$. Such mixing often occurs in family symmetry models through higher dimension operators involving familon fields beyond the leading order. 

In this paper we will explore both the charged lepton and neutrino origins for  $\theta_{13}$  in detail as well as the case where $\theta_{13}$ is partly driven by both sectors.  This is done for the cases that the unperturbed structure corresponds to TBM, GR and BM mixing. In Section \ref{secPMNS} we briefly review two parameterisations of the PMNS matrix that will be useful in our analysis. Section \ref{models} presents the TBM, GR and BM structures that provide the starting point for the perturbation analyses and Section \ref{family} discusses the family symmetry structure of these models. Sections \ref{perturbation} and \ref{secGR} discuss general neutrino mass perturbations about the symmetry limit that allow for non-zero $\theta_{13}$ for the case of approximate TBM, GR and BM mixing and derive the form of the Dirac CP violating phase and the mixing angles that result.   In Section \ref{charged} we give our assumptions about the mixing angles that diagonalize the charged lepton mass matrix. Section \ref{alternative} discusses how the results obtained still apply in cases that both the solar neutrino mass and the mass mixing (which are of similar magnitude) are generated as perturbations. In Section \ref{data} we provide a quantitative estimate for the CP violating phase and constraints on mixing angles, derived using the fits to the remaining mixing angles at both the  $1 \sigma$ and $3 \sigma$ levels of accuracy. This allows us to estimate the accuracy that will be needed to distinguish between the various cases.  In Section \ref{general} we discuss the case that the most general $3 \times 3$ mass mixing perturbations are allowed and argue that, if the solar mixing angle is not to receive unacceptable large corrections without fine tuning, such cases are disfavoured and the simpler $2 \times 2$ mixing case is preferred. Finally in Section \ref{summary} we present our Summary  and Conclusions.

\section{The PMNS matrix}\label{secPMNS}

The PMNS matrix arises from diagonalising the charged lepton and neutrino mass matrices, and involves the unitary matrices that transform the left-handed charged leptons and the left-handed neutrinos.  These unitary matrices in general each contain 3 Euler angles and 6 phases and can be put in the form
\bea
V^l&=& U_{23}^l U_{13}^l U_{12}^l P^l\nonumber\\
V^\nu &=& U_{23}^\nu U_{13}^\nu U_{12}^\nu P^\nu.
\label{vnu}
\eea
$P^{l}$ is a diagonal phase matrix that can be immediately removed by rephasing the charged lepton fields.  $P^\nu$ is a diagonal phase matrix $(e^{i \gamma_1}, e^{i \gamma_2}, e^{i \gamma_3})$ that can only be removed if the neutrinos are Dirac.    Each matrix $U^{l,\nu}_{ij}$ is itself unitary and contains two parameters:  a phase $\delta^{l,\nu}_{ij}$ and a rotation by angle $\theta^{l,\nu}_{ij}$ in the $ij$ plane 
\be
U^{l,\nu}_{ij} = \left (  \begin{array}{cc}
c^{l,\nu}_{ij} & s^{l, \nu}_{ij} e^{-i \delta^{l,\nu}_{ij}}  \\
 -s^{l, \nu}_{ij} e^{i \delta^{l,\nu}_{ij}} & c^{l,\nu}_{ij}  \end{array} \right)
\ee
and acts trivially in the direction orthogonal to the $ij$ plane.  
The resulting PMNS matrix is
\be
U_{PMNS}=V^{l \dagger} V^\nu
\ee
which involves 6 Euler angles and 12 phases.  

On the other hand the PMNS matrix can be written in the PDG form
\begin{eqnarray}\label{Eq:StandardParametrization}
 U_{PMNS}
 = \left(
  \begin{array}{ccc}
  c_{12}c_{13} &
  s_{12}c_{13} & s_{13}e^{-\I\delta}\\
  -c_{23}s_{12}-s_{13}s_{23}c_{12}e^{\I\delta} &
  c_{23}c_{12}-s_{13}s_{23}s_{12}e^{\I\delta}  &
  s_{23}c_{13}\\
  s_{23}s_{12}-s_{13}c_{23}c_{12}e^{\I\delta} &
  -s_{23}c_{12}-s_{13}c_{23}s_{12}e^{\I\delta} &
  c_{23}c_{13}
  \end{array}
  \right) \, P \, ,
  \label{PMNS}
\end{eqnarray}
where $s_{13}=\sin \theta_{13}$, $c_{13}=\cos \theta_{13}$
with $\theta_{13}$ being the
reactor angle, $s_{12}=\sin \theta_{12}$, $c_{12}=\cos \theta_{12}$
with $\theta_{12}$ being the
solar angle, $s_{23}=\sin \theta_{23}$, $c_{23}=\cos \theta_{23}$
with $\theta_{23}$ being the
atmospheric angle, $\delta$ is the (Dirac) CP violating phase which is in
principle measurable in neutrino oscillation experiments, and
$P = \mathrm{diag}(e^{\I \tfrac{\beta_1}{2}},
e^{\I\tfrac{\beta_2}{2}}, 0)$ contains additional (Majorana)
CP violating phases $\beta_1, \beta_2$.
Note that this form involves only 6 physical parameters, ($s_{ij}, \delta, \beta_1, \beta_2$)
 and corresponding to the parameterisation 
\be
U_{PMNS} = R_{23}U_{13}R_{12}P,
\ee
where $R_{ij}$ is a rotation in the $ij$ plane by angle $\theta_{ij}$.  

In general the relation between the parameters of these two representations is cumbersome, but to leading order in $s^e_{ij}$ and $s^\nu_{13}$ it simplifies \cite{Ohlsson:2005js,Antusch:2005kw,King:2002nf}
\bea
s_{13}e^{-i\delta_{13}}&=&s_{13}^{\nu}e^{-i\delta^{\nu}_{13}}-s_{12}^{l}s_{23}^{\nu}e^{-i(\delta_{23}^{\nu}+\delta_{12}^{l})}-s_{13}^{l}c_{23}^{\nu}e^{-i\delta_{13}^{l}} \label{s13}\nonumber \\
s_{12}e^{-i\delta_{12}}&=&s_{12}^{\nu}e^{-i\delta_{12}^{\nu}}-s_{12}^{l}c_{12}^{\nu}c_{23}^{\nu}e^{-i\delta_{12}^{l}}+s_{13}^{l}c_{12}^{\nu}s_{23}^{\nu}e^{i(\delta_{23}^{\nu}-\delta_{13}^{l})} \label{s12} \nonumber\\
s_{23}e^{-i\delta_{23}}&=&s_{23}^{\nu}e^{-i\delta_{23}^{\nu}}-s_{23}^{l}c_{23}^{\nu}e^{-i\delta_{23}^{l}} \label{s23}
\eea
where
\bea
\delta_{13} &=&\delta+\beta'_{1} \nonumber\\
\delta_{23} &=&\beta'_{2}\nonumber\\
\delta_{12} &=&\beta'_{1}-\beta'_{2}
\label{phases}
\eea
and 
\bea
\beta'_{1} &=& \beta_1 + (\gamma_1 - \gamma_3) \nonumber\\
\beta'_{2} &=& \beta_2 + (\gamma_2 - \gamma_3).
\label{phases2}
\eea
Results for $s_{ij}$ are obtained by taking the absolute values of eqs (\ref{s23}), while $\delta = \delta_{13} - \delta_{12} - \delta_{23}$.  Note that the phases $\gamma_i$ of $P^\nu$ lead to shifts in the Majorana phases $\beta_{1,2}$ but do not affect $\delta$ or $s_{ij}$.

\section{Models of lepton mixing}\label{models}
In this paper we only consider models in which diagonalisation of the neutrino mass matrix gives bi-maximal atmospheric mixing and $\theta_{13}=0$ at leading order. This class of models has a PMNS matrix of the form
\begin{eqnarray}
U_\Theta =
\left( \begin{array}{rrr}
\cos\Theta & \sin \Theta &0\\
\frac{-\sin \Theta}{\sqrt{2}} & \frac{\cos \Theta}{ \sqrt{2}} & \frac{1}{\sqrt{2}}\\
\frac{\sin \Theta}{\sqrt{2}} & \frac{-\cos \Theta}{ \sqrt{2}} & \frac{1}{\sqrt{2}}
\end{array}
\right)P
\label{GEN}
\end{eqnarray}
and the phase ambiguity discussed above allows us to adopt a particular phase convention in which all mixing angles are in the first quadrant.  This form can arise from a $Z_2 \times Z_2$ symmetry \cite{Ge:2011ih, Ge:2011qn}, as discussed later.  The associated Lagrangian for the neutrino masses is given by
\be
{\cal L}=\frac{1}{2} ( m_3 \, \nu_a^2+m_2 \, \nu_b^2 + m_1 \, \nu_c^2 ) \; + h.c.
\label{L}
\ee
where the mass eigenstates are given by $\nu_{a,b,c} \equiv (n_{a,b,c} \cdot \nu)$ which point in the particular directions $n_{a,b,c}$ in flavour space given by
\bea
\nu_{a} &\equiv& (\nu_{\mu}+\nu_{\tau})/\sqrt{2} \nonumber\\
\nu_{b} &\equiv& s_{\Theta}\nu_{e}+c_{\Theta}(\nu_{\mu}-\nu_{\tau})/\sqrt{2} \nonumber\\
\nu_{c} &\equiv& c_{\Theta}\nu_{e}-s_{\Theta}(\nu_{\mu}-\nu_{\tau})/\sqrt{2}. 
\label{nudir}
\eea
The origin of the $Z_2 \times Z_2$ symmetry follows from the fact that eq(\ref{L}) is invariant under $\nu_i\rightarrow -\nu_i$. The case that there is an overall sign change of all three states is irrelevant leaving just the $Z_2 \times Z_2$ symmetry. In Section \ref{TBMperturbations} we perturb the system by adding to the Lagrangian a bilinear term of the form $\nu_i\nu_j$ which breaks the symmetry to a single $Z_2$.

The above structure, and the entire analysis of this paper, applies to both the normal and inverted hierarchies. For the normal hierarchy $m_3 > m_{1,2}$ while for the inverted hierarchy $m_3 < m_{1,2}$.  The atmospheric oscillation is governed by the large mass splitting between $m_3$ and $m_{1,2}$, while the solar oscillation is governed by the smaller splitting between $m_2$ and $m_1$.  In both normal and inverted hierarchies the lightest neutrino can be massless.

\subsection{Pure Tri-Bi-Maximal mixing}

Pure TBM mixing \cite{Harrison:2002er,Harrison:2002kp} is generated by a PMNS matrix of the form of eq(\ref{GEN}) with $\tan\Theta=1/\sqrt{2}$, i.e.
\begin{eqnarray}
U_{TBM} =
\left( \begin{array}{rrr}
\sqrt{\frac{2}{3}}  & \frac{1}{\sqrt{3}} & 0 \\
-\frac{1}{\sqrt{6}}  & \frac{1}{\sqrt{3}} & \frac{1}{\sqrt{2}} \\
\frac{1}{\sqrt{6}}  & -\frac{1}{\sqrt{3}} & \frac{1}{\sqrt{2}}
\end{array}
\right)P\equiv R_{23}R_{12}P
\label{MNS0}
\end{eqnarray}
where $R_{ij}$ are the orthogonal matrices corresponding to the mixing angles, $s_{12}^{2}=1/3$, $s_{23}^{2}=1/2$ and $s_{13}=0$.
Assuming the charged lepton mass matrix is diagonal this form can arise from the neutrino mass matrix if there is a suitable non-Abelian family symmetry. In Section \ref{perturbation} we discuss how this matrix changes if one allows small perturbations in the neutrino sector about the TBM form. However in general we expect there also to be corrections to the TBM form coming from the diagonalisation of the charged lepton mass matrix and these must be included via eqs(\ref{s23}).

The structure of $V^{\nu}$  depends on the basis chosen for the states. In models of TBM mixing that follow from an underlying family symmetry one starts from the symmetry basis in which matter states belong to unmixed representations of the family group. In this basis we begin by taking $V^l=1$ so that the neutrino current eigenstates are $\nu_{e,\mu,\tau}$; corrections from $V^l\neq 1$ are discussed later.  Then the pure TBM Lagrangian describing the neutrino masses has the form of eq(\ref{L}) where
the mass eigenstates are given by eq(\ref{nudir}) with $\tan\Theta=1/\sqrt{2}$, i.e.
\bea
\nu_{a} &\equiv& (\nu_{\mu}+\nu_{\tau})/\sqrt{2} \label{nudirTBMa}\nonumber\\
\nu_{b} &\equiv& (\nu_{e}+\nu_{\mu}-\nu_{\tau})/\sqrt{3} \label{nudirTBMb}\nonumber \\
\nu_{c} &\equiv& (2\nu_{e}-\nu_{\mu}+\nu_{\tau})/\sqrt{6}. \label{nudirTBMc}
\eea
In the absence of any perturbations we have pure TBM. The columns of $U_{TBM}$ are given by the direction vectors $n_{c,b,a}$.
In the notation of Section \ref{secPMNS}, pure TBM corresponds to the case $V^l=1$ and $U_{13}^\nu=1$ in eq(\ref{vnu}). The combination $U_{23}^\nu U_{13}^\nu U_{12}^\nu$ can then be rewritten as $P^{l'}R_{23}R_{12}P^{\nu'}$, the phases in $P^{l'}$ can be absorbed in a redefinition of the charged lepton fields and those in $P^{\nu'}$ can be absorbed in $P^{\nu}$. Thus one obtains $V^\nu=R_{23}R_{12}P^{\nu}$ giving the TBM of eq(\ref{MNS0}).

\subsection{Golden Ratio mixing}
Another promising form for the CKM matrix that can be obtained from an underlying family symmetry is the Golden Ratio case \cite{Datta:2003qg,Kajiyama:2007gx} given by eq(\ref{GEN}) with
$\tan\Theta=1/\phi$ where  $\phi$ is the Golden Ratio $(1+\sqrt{5})/2$. This leads to the mixing angle $\theta_{12}$  given by $s_{12}^2=1/(\sqrt{5}\phi)\approx 0.276$, quite close to the fitted value. An alternative version of GR mixing\cite{Rodejohann:2008ir,Adulpravitchai:2009bg}
 has been proposed in which $\tan\Theta=2/\phi$ corresponding to $\sin^2\theta_{12}=0.345$, again quite close to the fitted value.
In both cases the mass eigenstates in eq(\ref{L}) are given by eq(\ref{nudir}).

\subsection{Bi-maximal mixing}
For comparison we will also consider  bi-maximal mixing \cite{Fukugita:1998vn,Barger:1998ta,Davidson:1998bi} with the neutrino contribution to the CKM matrix given by eq(\ref{GEN}) with $\tan\Theta=1$ corresponding to 
\begin{eqnarray}
V^\nu_{BM} =
\left( \begin{array}{rrr}
\frac{1}{\sqrt{2}} &\frac{1}{\sqrt{2}} & 0 \\
-\frac{1}{2}  & \frac{1}{2} & \frac{1}{\sqrt{2}} \\
\frac{1}{2}  &- \frac{1}{2} & \frac{1}{\sqrt{2}}
\end{array}
\right)
\label{BM}
\end{eqnarray}
giving $s_{12}^{\nu 2}=s_{23}^{\nu 2}=0.5$ and $s_{13}=0$, the reality of the matrix following from the underlying family symmetry. In this case an acceptable value for $\theta_{12}$ requires large corrections from the charged lepton sector with, analogous to the quark sector, sizeable mixing in the (1,2) sector. Choosing
\begin{eqnarray}
V^l_{ BM} =
\left( \begin{array}{rrr}
\cos\alpha&-e^{-i\delta^l}\sin\alpha & 0 \\
e^{-i\delta^l}\sin\alpha  &\cos\alpha & 0 \\
0& 0&1
\end{array}
\right)
\label{BML}
\end{eqnarray}
one finds 
\bea
s_{13}&\approx&\alpha/\sqrt{2}\nonumber\\
s_{12}^2&\approx&1/2+\alpha\cos\delta/\sqrt{2}\nonumber\\
s_{23}^2&\approx&1/2-\alpha^2/4\nonumber\\
\delta&=&\delta^l.
\label{BMM}
\eea
\section{Family symmetry structure}\label{family}

In this section we discuss the discrete symmetry structure that leads to the unperturbed TBM, GR and BM schemes in the neutrino sector. For these cases the neutrino mass matrix has a $Z_2 \otimes Z_2$ family symmetry. One $Z_2$ is the same for all three schemes and is the $\mu \leftrightarrow -\tau$ interchange symmetry, with generator 
\begin{eqnarray}
U
 = \left(
\begin{array}{ccc}
 1 &0 &0\\
  0&0  &-1\\
  0 &-1&0
  \end{array} 
  \right).
  \label{T}
\end{eqnarray}

$U$ has three eigenvectors given by the $\nu_{a,b,c}$ of eq(\ref{nudir}) with eigenvalues $-1,+1,+1$ respectively and so the mass matrix invariant under $U$ also has these three eigenvectors.  The eigenvector with negative eigenvalue is $\nu_a = (0,1,1)/\sqrt{2}$ and leads to bimaximal atmospheric neutrino mixing in all three schemes. The other two eigenstates have positive eigenvalues and can mix with an undetermined angle $\theta_{12}$, c.f. eq(\ref{PMNS}).  Since $\nu_a$ is prevented from mixing with these two states,  $U$ is also responsible for setting  $\theta_{13}=0$.   

The remaining $Z_2$ symmetry, with generator $S$, also has the states $\nu_a,\;\nu_b$ and $\nu_c$ of eq(\ref{nudir}) as eigenvectors:  $S \, \nu_{a,b,c} = P_{a,b,c} \, \nu_{a,b,c}$. The eigenvalues $P_b$ and $P_c$ have opposite signs, $P_bP_c=-1$,  preventing mixing between $\nu_b$ and $\nu_c$.    The general form for $S$ for arbitrary $\Theta$ is
 \begin{eqnarray}
S
 = \frac{P_a}{\beta-\gamma} \left(
    \begin{array}{ccc}
 1 &\alpha&-\alpha \\
\alpha&\beta  &-\gamma\\
  -\alpha &-\gamma&\beta
  \end{array}
  \right)
  \label{S}
\end{eqnarray}
where 
\bea
\alpha&=&\frac{\sqrt{2}}{\tan\Theta-\cot\Theta}\nonumber\\
\beta&=&\frac{(P_aP_b+1)\cot\Theta+(P_aP_b-1)\tan\Theta}{2(\tan\Theta-\cot\Theta)}\nonumber\\
\gamma&=&\frac{-(P_aP_b-1)\cot\Theta-(P_aP_b+1)\tan\Theta}{2(\tan\Theta-\cot\Theta)}.
\label{GRS}
\eea
A particular form for $S$, with a fixed value of $\Theta$, determines $\theta_{12}$.

Models built to generate these various mixing schemes are of two types (for reviews see \cite{King:2013eh,Altarelli:2010gt}). The first type has a family symmetry, $G_f$,  that is broken to $G_{\nu}=Z_2 \otimes Z_2$ in the neutrino sector.   
The second type has a family symmetry that does not contain $S$ and $U$ but has flavon fields, $\phi_{a,b}$, that, due to the family symmetry, acquire vevs pointing in the $n_{a,b}$ directions, $\phi_{a,b} \propto n_{a,b}$. If these fields have opposite charge under a $Z_{2}'$ non-family-symmetry the mass eigenstates then are given by $\nu_{a,b}=\phi_{a,b} \cdot \nu$. The Lagrangian then has the form given in  eq(\ref{L}) and is invariant under $S$ and $U$ corresponding to an emergent  $Z_2 \otimes Z_2$ symmetry. An advantage of this approach is that the underlying family symmetry readily applies to the quark sector too.

\subsection{Tri-Bi-Maximal mixing}\label{SectionTBM}
TBM can be generated by an underlying $S_4$, which contains both $S$ and $U$, or by $A_4$ which contains $S$ and, for a restricted non-family-symmetry breaking pattern, leads to a Lagrangian with an accidental symmetry generated by $U$\footnote {See \cite{King:2013eh,Altarelli:2010gt} and references therein.}.
For the TBM case  S, for $P_{a}P_{b}=+1$, has the form given by eq(\ref{S}) with $\tan\Theta=1/\sqrt{2}$
 \begin{eqnarray} 
S_{TBM}
 =  \frac{P_{a}}{3} \left(
    \begin{array}{ccc}
 -1 &2 &-2 \\
  2&2  &1\\
  -2 &1&2
  \end{array}
  \right). 
\end{eqnarray}
Its eigenvectors $\nu_{a,b,c}$ have eigenvalues $P_{a}(+1,+1,-1)$ respectively.  Thus, while the symmetry generated by both $U$ and $S_{TBM}$ is unbroken, there cannot be mixing between the states and the mass eigenstates are just $\nu_{a,b,c}$  corresponding to TBM mixing.

\subsection{Golden ratio mixing}
The first version of GR mixing, which we label GR1, has $\tan\Theta=1/\phi$ and can be obtained from an underlying $A_5$ family symmetry\cite{Feruglio:2011qq}; while the second version of GR mixing, which we label GR2, has $\tan\Theta=2/\phi$ and can be obtained from the dihedral group $D_{10}$ \cite{Adulpravitchai:2009bg,Blum:2007jz}.

In the neutrino sector, the symmetry is broken to $G_{\nu}=Z_2 \otimes Z_2$ where the first $Z_2$ is generated by the $\mu\leftrightarrow \tau$ interchange generator $U$ of eq(\ref{T}) with eigenvalues $+1,-1,-1$. The second $Z_2$ has the generator, S, given by eq(\ref{S}).

For GR1 the generator with $P_aP_b=1$ has a relatively simple form
\begin{eqnarray}
S_{GR}
 = \frac{P_{a}}{\sqrt{5}} \left(
    \begin{array}{ccc}
 -1 &\sqrt{2} &-\sqrt{2} \\
  \sqrt{2}&\phi  &1/\phi\\
 - \sqrt{2} &1/\phi&\phi
  \end{array}
  \right)
  \label{S1}
\end{eqnarray}
leading to $(P_a,P_b,P_c)= P_{a}(+1,+1,-1)$.
The case with  $P_aP_b=-1$ in eq(\ref{GRS}) is given by $S_{GR}U$ with $(P_a,P_b,P_c)= P_{a}(-1,+1,-1)$.

In the case of GR2 the generator with  $P_aP_b=1$ is more complicated 
\begin{eqnarray}
S_{GR}
 = \frac{P_{a}}{2/\phi+\phi/2} \left(
    \begin{array}{ccc}
 2/\phi-\phi/2 &\sqrt{2} &-\sqrt{2} \\
  \sqrt{2}&\phi/2  &2/\phi\\
 - \sqrt{2} &2/\phi&\phi/2
  \end{array}
  \right)
  \label{S1}
\end{eqnarray}
making it somewhat less attractive.

\subsection{Bi-maximal mixing}
Bi-maximal mixing can be generated by an underlying $S_4$ symmetry \cite{Altarelli:2009gn} broken to $G_{\nu}=Z_2 \otimes Z_2$. As above  BM mixing in the atmospheric neutrino sector and the vanishing of $\theta_{13}^{\nu}$ is driven by the first $Z_2$ factor with the generator $U$ of eq(\ref{T}). The vanishing of the CP violating phase and maximal mixing for $s_{12}$  is due to the second $Z_2$ factor with generator given by eq(\ref{S}) with $\tan\Theta=1$
\begin{eqnarray}
S_{BM}
 =\frac{1}{2} \left(
    \begin{array}{ccc}
 0 &\sqrt{2} &-\sqrt{2} \\
  \sqrt{2}&1  &1\\
  -\sqrt{2} &1&1
  \end{array}
  \right)
  \label{S2}
\end{eqnarray}
with eigenvectors $\nu_{a,b,c}$ given by eq(\ref{nudir}) with $\tan\Theta=1$ and eigenvalues $+1,+1,-1$ respectively.

\section{Perturbations to TBM from the Neutrino Sector}\label{perturbation}\label{TBMperturbations}

In TBM mixing the neutrino directions $\nu_{a,b,c}$ of eq(\ref{nudirTBMc}) play a special role and we therefore parameterize deviations from pure TBM by allowing for small mixing between these states.  There are three independent perturbations about eq(\ref{MNS0}) arising from mass terms $\nu_a \nu_b, \;\nu_a \nu_c$ and  $\nu_b \nu_c$ and as a result the mass eigenstates are mixtures of the original eigenstates. In leading order this corresponds to the parameterisation given by
\begin{eqnarray}
U
 \approx \left(
    \begin{array}{ccc}
\frac{2} {\sqrt{6}} (1-t^*)&\frac{1}{\sqrt{3}}(1+2t) &\frac{1}{\sqrt{2}}(r+2s) \\
  -\frac{1}{\sqrt{6}}(1+3s^*+2t^*)&\frac{1}{\sqrt{3}}(1-\frac{3}{2}r^*-t) &\frac{1}{\sqrt{2}}(1+r-s)\\
  \frac{1}{\sqrt{6}}(1-3s^*+2t^*)&-\frac{1}{\sqrt{3}}(1+\frac{3}{2}r^*-t)&\frac{1}{\sqrt{2}}(1-r+s)
  \end{array}
  \label{mixing}
  \right)
\end{eqnarray}
where $r,s$ and $t$ are small complex constants that quantify the effect of $\nu_a \nu_b, \;\nu_a \nu_c$ and  $\nu_b \nu_c$ mixing respectively\footnote{A related expansion was given in \cite{King:2007pr}}.

These three perturbations correspond to a (small) breaking of the underlying $Z_2\otimes Z_2$ symmetry to a $Z_2$ subgroup generated by $S_{TBM}$, $S_{TBM}U$ and $U$ respectively. To generate a non-zero  $\theta_{13}$ the symmetry generated by $U$ must be broken and so we do not consider the case of $\nu_{b}\nu_{c}$ mixing. The residual symmetries $S_{TBM}$ and $S_{TBM}U$ prevent $\nu_{b}\nu_{c}$ mixing and thus  keep small the deviation of $\theta_{12}$ from the tri-maximal form while allowing for a viable value of $\theta_{13}$.

In an $A_4$ theory $\nu_{a}\nu_{b}$ mixing is natural in the sense that the $U$ symmetry is an accidental symmetry and can be broken by an $A_4$ singlet family vev, leaving the symmetry $S_{TBM}$ in $A_{4}$ unbroken. In $S_4$ the $U$ symmetry can be spontaneously broken by an $S_4$ non-singlet  family field but requires a mechanism to generate the appropriate vacuum alignment.   In both $A_4$ and $S_4$ cases, the breaking that leaves a $Z_2$ symmetry generated by  $S_{TBM}U$, corresponding to $\nu_{a,c}$ mixing, requires a suitable vacuum alignment of a familon vev. 

For the case the symmetry is emergent the $\nu_a \nu_b$ mixing is natural in the sense that it occurs if the $Z_2'$ non-family-symmetry, that prevents such mixing, is broken by higher dimension operators, perturbations to the vevs of $\phi_{a,b}$ or from vevs of additional family-symmetry-singlet flavon fields.  The generation of  $\nu_b \nu_c$ or  $\nu_a \nu_c$ mixing requires the addition of a $\phi_c$ familon vev.

In the rest of this Section we assume that a single perturbation dominates. The cases $\nu_a \nu_b$ and $\nu_a \nu_c$ mixing, equivalent to $r\ne 0$ and $s\ne 0$ respectively, correspond to TM1 and TM2 mixing discussed in \cite{Albright:2008rp} and to the $p=+1$ and $p=-1$ schemes of  \cite{Ge:2011qn}.  They lead to relations between phases and mixing angles that are discussed below. 
We discuss the case of more general mixing in section \ref{general} where we demonstrate that, although it is possible to have more than a single mass-mixing operator, if the correction to tri-maximal mixing in the solar neutrino sector is to be naturally small, it is likely that a single mass-mixing operator should dominate

\subsection{$\nu_{a}\nu_{b}$ mixing:  $Z_2 \otimes Z_2\rightarrow S_{TBM}$}
The Lagrangian describing neutrino mass is modified to have the form
\be
{\cal L}_{ab} = \frac{1}{2} \Big( m_3(\nu_{a}+\sqrt{3}\epsilon\nu_{b})^{2}+m_2 (\nu_{b}-\sqrt{3}\epsilon^{*}\nu_{a})^{2}+m_1\nu_{c}^{2} \Big) \; + h.c.
\label{ab}
\ee
where $\epsilon$ is a (small) complex expansion parameter.  This case corresponds to the choice $r= \sqrt{2} \epsilon$ and $s=t=0$ in eq(\ref{mixing}).  Writing $\epsilon=e^{-i\delta^{\epsilon}}\s_{13}^{\epsilon}$ it is straightforward to construct $V_{\nu_{L}}$. Comparing with eq(\ref{PMNS}) one finds 

\bea
\delta_{23}^{\nu}&\approx&2\sqrt{2}s_{13}\sin\delta^{\epsilon}+\pi\nonumber\\
\delta_{13}^{\nu}&\approx&\delta^{\epsilon}+\sqrt{2}s_{13}\sin\delta^{\epsilon}+\pi\nonumber\\
\delta_{12}^{\nu}&\approx&-\sqrt{2}s_{13}\sin\delta^{\epsilon}\nonumber\\
\s_{13}^{\nu}&\approx&\s_{13}^{\epsilon}\nonumber\\
\s_{23}^{\nu}&\approx&|1/\sqrt{2}+e^{-i\delta^{\epsilon}}\s^{\epsilon}_{13}|\nonumber\\
\s_{12}^{\nu}&\approx&1/\sqrt{3}.
\label{nuab}
\eea
Hence the Dirac neutrino phase is given by $\delta^{\nu}\equiv \delta_{13}^{\nu}- \delta_{23}^{\nu}- \delta_{12}^{\nu}=\delta^{\epsilon}$. Note that the initial unperturbed TBM form taken here is as given in eq(\ref{MNS0}) and that this contains two unknown phases. This means that the Majorana phases are not determined and this conclusion also applies to all the mixing schemes considered in this paper.

\subsection{$\nu_{a}\nu_{c}$ mixing:  $Z_2 \otimes Z_2\rightarrow S_{TBM}U$}
The modified Lagrangian now has the form
\be
{\cal L}_{ac}= \frac{1}{2} \Big( m_3 (\nu_{a}+\sqrt{\frac{3}{2}}\epsilon\nu_{c})^{2}+m_2 \, \nu_{b}^{2}+m_1(\nu_{c}-\sqrt{\frac{3}{2}}\epsilon^{*}\nu_{a})^{2} \Big)  \; + h.c.
\ee
corresponding to the choice $s=  \epsilon /  \sqrt{2}$ and $r=t=0$ in eq(\ref{mixing}).
Once again with the definition  $\epsilon=e^{-i\delta^{\epsilon}}\s_{13}^{\epsilon}$ and comparing with eq(\ref{vnu}) and extracting the Dirac phase one finds 

\bea
\s_{13}^{\nu}&\approx&\s_{13}^{\epsilon}\nonumber\\
\s_{23}^{\nu}&\approx&|1/\sqrt{2}-e^{-i\delta^{\epsilon}}\s^{\epsilon}_{13}/2|\nonumber\\
\s_{12}^{\nu}&\approx&1/\sqrt{3}\nonumber\\
\delta^{\nu}&=&\delta^{\epsilon}.
\label{nuac}
\eea

\section{Perturbations to GR from the Neutrino Sector}\label{secGR}
 As for the TBM case a non-zero value for $s_{13}$ requires that the $Z_2$ factor generated by $U$ be broken. To prevent large deviations of $s_{12}$ from the GR value another $Z_2$ factor should remain. For $\nu_a\nu_b$ mixing $P_aP_b=1$ in eq(\ref{GRS}), which we call $S_{GR}$, while for $\nu_a\nu_c$ mixing $P_aP_b=-1$ corresponding to $S_{GR}U$.
 
\subsection{$\nu_{a}\nu_{b}$ mixing:  $Z_2 \otimes Z_2\rightarrow S_{GR}$}
The Lagrangian describing neutrino mass is given by 
\be
{\cal L}_{ab} = \frac{1}{2} \Big( m_3(\nu_{a}+\frac{\epsilon}{\sin\Theta}\nu_{b})^{2}+m_2 (\nu_{b}-\frac{\epsilon^*}{\sin\Theta}\nu_{a})^{2}+m_1 \, \nu_{c}^{2} \Big)  \; + h.c.
\label{abgr}
\ee
 Comparing with eq(\ref{vnu}) one finds 
\bea
\s_{13}^{\nu}&\approx&\s_{13}^{\epsilon}\nonumber\\
\s_{23}^{\nu}&\approx&\frac{1}{\sqrt{2}}|1+e^{-i\delta^{\epsilon}}\cot\Theta\s^{\epsilon}_{13}|\nonumber\\
\s_{12}^{\nu}&\approx&s_{\Theta}\nonumber\\
\delta^{\nu}&=&\delta^{\epsilon}
\label{nuabgr}
\eea
where, as always, we define $\epsilon=e^{-i\delta^{\epsilon}}\s_{13}^{\epsilon}$.

\subsection{$\nu_{a}\nu_{c}$ mixing:  $Z_2 \otimes Z_2\rightarrow S_{GR} U$}
In this case we have
\be
{\cal L}_{ac}= \frac{1}{2} \Big( m_3 (\nu_{a}+\frac{\epsilon}{c_{\Theta}}\nu_{c})^{2}+m_2 \, \nu_{b}^{2}+m_1(\nu_{c}+\frac{\epsilon^*}{c_{\Theta}}\nu_{a})^{2} \Big)  \; + h.c.
\ee
giving
\bea\s_{13}^{\nu}&\approx&\s_{13}^{\epsilon}\nonumber\\
\s_{23}^{\nu}&\approx&\frac{1}{\sqrt{2}}|1-e^{-i\delta^{\epsilon}}\tan\Theta\s^{\epsilon}_{13}|\nonumber\\
\s_{12}^{\nu}&\approx&s_{\Theta}\nonumber\\
\delta^{\nu}&=&\delta^{\epsilon}.
\label{nuacgr}
\eea

\section{The charged lepton sector}\label{charged}

In order to compare TBM and GR perturbation predictions with experiment it is necessary to determine the contribution from the charged lepton sector. In this we are partly guided by the relations between down quark and charged lepton masses implied by an underlying GUT and we assume that the charged lepton mass matrix is hierarchical, with a similar structure to the quark mass matrices.  

Given the smallness of the equivalent quark mixing, we set $s^l_{13} = 0$ and drop the last term in eqs(\ref{s13}). Similarly we choose $s_{23}^{l}=O(\frac{m_{\mu}}{m_{\tau}})$ in analogy with the equivalent quark mixing angle, $V^{CKM}_{13}=O(m_{s}/m_{b})$.
For the case of $\theta_{12}^{l}$ we will illustrate the possibilities by three choices. The first choice assumes the simplest possibility, namely $\theta_{12}^{l}=0$. The second case, arguably the most plausible as it is the analogue of the successful relation in the down quark sector discussed above, is given by $\s_{12}^{l}=\sqrt{\frac{m_{e}}{m_{\mu}}}$. Finally, $\theta_{13}$ may arise entirely from the charged lepton contribution, with $s_{12}^{l} \approx \sqrt{2} \, s_{13}$ \cite{Antusch:2011qg,Antusch:2012fb,Marzocca:2011dh,leptonmixing,Xing:2001cx}. For example, this can result if the  renormalisable (dimension 4) Yukawa couplings are forbidden by a symmetry and the coupling is generated by a higher dimension 5 term \cite{Antusch:2009gu}. Although in this case  the neutrino mass matrix is not perturbed away from TBM we include it for comparison with the perturbed cases.  Note that although It is usual to compute $\delta$ treating $s_{12}^l$ as a perturbation, in this case it introduces significant errors so in Section \ref{data} we will keep the full expression.

\section{An alternative perturbative framework}\label{alternative}
In the previous sections we have described the well-known TBM and GR mixing schemes, as well as their origin from family symmetries and perturbations about them that yield non-zero $\theta_{13}$.  In these schemes there is no connection between the masses and mixing angles. Yet, for the normal hierarchy, the observed value for $\theta_{13}$ suggests an alternative scheme where only one neutrino mass arises at leading order.   Let $\nu_{a,b,c}$ be some special directions in flavor space such that $\theta_{13}$ arises from $ab \, (ac)$ mixing, so that the three largest neutrino mass operators are
\be
{\cal L}=\frac{1}{2} ( m_3 \, \nu_a^2+m_2 \, \nu_b^2 ) + m_{ab} \, \nu_a \nu_b \, (m_{ac} \, \nu_a \nu_c)  \; + h.c..
\label{alt}
\ee
The relative sizes of these mass terms are $m_3:m_2:m_{ab} \, (m_{ac}) = 1: 0.16:0.27 \,(0.19)$.  Remarkably we see that the perturbation that induces mixing is {\it larger} than the solar mass term $m_2$.  This motivates searching for a new type of theory where only the atmospheric neutrino mass arrives at leading order, ${\cal L}_0=m_3 \, \nu_a^2/2$, preserving a $U(2)$ symmetry, while the perturbation includes both solar and mixing terms, 
${\cal L}'= m_2 \, \nu_b^2/2  + m_{ab} \, \nu_a \nu_b \, (m_{ac} \, \nu_a \nu_c)$.
 
As a very simple example consider a theory where the neutrino fields $\nu$ and two flavon fields $\phi_{a,b}$ transform under a flavor group $G_f$, and there is an additional $Z_3$ symmetry under which $\nu$ and $\phi_a$ are singlets, but $\phi_b$ transforms as $\alpha$, with $\alpha^3=1$.  In addition there is  a flavor singlet field $\chi$ with $Z_3$ transformation $\alpha$ that acquires a vev much less than the cutoff, giving a small dimensionless parameter $\epsilon$ that transforms as $\alpha$.   The flavon fields acquire vevs of magnitudes $v_{a,b}$ in special directions $n_{a,b}$ in flavor space, $\phi_{a,b} = v_{a,b} n_{a,b}$, that could be the directions of  eqns(\ref{nudirTBMc}) for example. This leads to neutrinos $\nu_{a,b} = (\phi_{a,b} \cdot \nu)$ transforming as $(1,\alpha)$ under $Z_3$.  The Lagrangian for neutrino masses is then a perturbation series in the small field $\epsilon = \chi / \Lambda $
\be
{\cal L}_{\nu}= m_\nu  \left( \frac{1}{2} c_1 \, \nu_a^2   +  \epsilon \frac{1}{2} c_2 \, \nu_b^2  + \epsilon^* c_3 \, \nu_a \nu_b + {\cal O} (\epsilon^2) \right)  \; + h.c.
\label{altpert}
\ee
where $m_\nu$ is the overall neutrino mass scale, $c_{1,2,3}$ are order unity dimensionless couplings, and $\Lambda$ is the UV cutoff.  The observed masses are accounted for by taking $\epsilon \sim 0.2$.
In practice the predictions for the mixing angles and the phases are the same as for the cases discussed above but the scheme has the merit of relating the magnitude of the solar neutrino mass to the magnitude of $\theta_{13}$ in terms of a single expansion parameter $\epsilon$.   This result would be spoiled if $v_a$ and $v_b$ are not comparable, but could be regained by having $m_b$ and $m_{ab}$ arise at the same order in $v_a$ and $v_b$, as occurs in the supersymmetric theory described by the superpotential 
\be
{\cal W}_{\nu}= m_\nu  \left( \frac{1}{2} c_1 \, \nu_a^2   +  \frac{1}{2} c_2 \, \frac{\chi \cdot \phi_a}{\Lambda^2} \; \nu_b^2  +  c_3 \, \frac{\chi \cdot \phi_b}{\Lambda^2} \; \nu_a \nu_b + ... \right)
\label{altpert}
\ee
where, for example, $\nu$ and $\chi$ are triplets under $G_f$ and $\phi_{a,b}$ are anti-triplets, and the $Z_3$ quantum numbers are as given above.
 
In the next section we give our predictions for neutrino mixing angles and CP violation when $ab$ and $ac$ neutrino mixing is added to the TBM and GR schemes.   A key point is that these predictions are more general, resulting whenever neutrino masses are dominated by the three operators $\nu_a \nu_a, \nu_b \nu_b, \nu_a \nu_b \, (\nu_a \nu_c)$ for the normal hierarchy, and by $\nu_b \nu_b, \nu_c \nu_c,  \nu_a \nu_b \, (\nu_a \nu_c)$ for the inverted hierarchy.  They are independent of the organization of the perturbation theory,  and depend only on the special directions $n_{a,b}$.    
 
\section{Comparison with data}
\label{data}

We organise our analysis according to the size of $s_{12}^{l}$: 0, $\sqrt{m_e/m_\mu}$ or $\sqrt{2} \, s_{13}$. In all cases we include the error that results from a charged lepton mixing angle given by $s_{23}^l=m_{\mu}/m_{\tau}$.

\begin{table}[t]
\caption{Predictions for the allowed ranges of the Dirac CP violating phase $\delta$ and the mixing angles.   Two columns are shown for the prediction of $\delta$ corresponding to the $1\sigma$  and $3\sigma$ ranges of $s_{23}$ and $s_{12}$ of the fit of Gonzales-Garcia et al \cite{GonzalezGarcia:2012sz}.  For comparison, the final 2 rows show the fit to $\delta$ and $s_{12}$ from data  (c.f. Table \ref{fits}). A dash for a mixing angle indicates that the fitted value has been used to determine the allowed range of the other parameters. A dash for the phase indicates no solutions possible.}
\centering
\begin{tabular}{|c|c|c||c|c|c|c|}
\hline 
Model&$\nu$ perturbation& $s_{12}^{l}$ & $\delta/\pi$ (1$\sigma$) & $\delta/\pi$ (3$\sigma$)&$s_{12}^2$&$s_{23}^2$\\
\hline \hline
TBM&$\nu_{ab}$ mixing (NH) & 0 & $\pm$(0.58--0.79) & 0 -- 2 &0.33 &-\\
 &(IH) & 0 & $\pm$(0.23--0.52)  & 0 -- 2 &0.33&- \\
&(NH)& $\sqrt{\frac{m_{e}}{m_{\mu}}}$& $\pm$(0.58--1) & 0 -- 2& 0.29--0.38&-\\
&(IH)& $\sqrt{\frac{m_{e}}{m_{\mu}}}$& $\pm$(0--0.54) & 0 -- 2& 0.29--0.38&-\\
\hline
TBM&$\nu_{ac}$ mixing (NH) & 0& $\pm$(0--0.38) & 0 -- 2 &0.33&- \\
&(IH) & 0& $\pm$(0.51--1) & 0 -- 2 &0.33&- \\
&(NH)&$\sqrt{\frac{m_{e}}{m_{\mu}}}$&$\pm$(0--0.4)  & 0 -- 2& 0.29--0.38&-\\
&(IH)&$\sqrt{\frac{m_{e}}{m_{\mu}}}$&$\pm$(0.5--1)  & 0 -- 2& 0.29--0.38&-\\
\hline
TBM&None& $\sqrt{2} \, s_{13}$& $\pm$(0.54--0.61) & $\pm$(0.48--0.68)&- &0.43--0.55 \\
\hline \hline
GR1&$\nu_{ab}$ mixing (NH) & 0 & $\pm$(0.58--0.76) &0 -- 2 &0.28 &-\\
& (IH) & 0 & $\pm$(0.27--0.53) &0 -- 2 &0.28 &-\\
&(NH)& $\sqrt{\frac{m_{e}}{m_{\mu}}}$& $\pm$(0.6--0.97) & 0 -- 2& 0.23--0.32&-\\
&(IH)& $\sqrt{\frac{m_{e}}{m_{\mu}}}$& $\pm$(0--0.54) & 0 -- 2& 0.23--0.32&-\\
\hline
GR1&$\nu_{ac}$ mixing (NH)& 0&$ \pm$(0--0.37)& 0 -- 2 &0.28 &-\\
GR1& (IH) & 0&  $\pm$(0.52--1)& 0 -- 2 &0.28 &-\\
&(NH)&$\sqrt{\frac{m_{e}}{m_{\mu}}}$& $\pm$(0--0.39) &0 -- 2& 0.23--0.32 &-\\
&(IH)&$\sqrt{\frac{m_{e}}{m_{\mu}}}$& $\pm$(0.5--1) &0 -- 2& 0.23--0.32 &-\\
\hline
GR1&None& $\sqrt{2} \, s_{13}$& $\pm$(0.41--0.47) & $\pm$(0.32--0.51)&-&0.43-0.55  \\
\hline \hline
GR2&$\nu_{ab}$ mixing (NH) & 0 & $\pm$(0.61--1) &0 -- 2 &0.35 &-\\
& (IH) & 0 & $\pm$(0--0.51) &0 -- 2 &0.35 &-\\
&(NH)& $\sqrt{\frac{m_{e}}{m_{\mu}}}$& $\pm$(0.66--1) & 0 -- 2& 0.31--0.40&-\\
&(IH)& $\sqrt{\frac{m_{e}}{m_{\mu}}}$& $\pm$(0--0.47) & 0 -- 2& 0.31--0.40&-\\
\hline
GR2&$\nu_{ac}$ mixing (NH)& 0&$ \pm$(0.16--0.41)& 0 -- 2 &0.35 &-\\
GR2& (IH) & 0&  $\pm$(0.48--0.87)& 0 -- 2 &0.35 &-\\
&(NH)&$\sqrt{\frac{m_{e}}{m_{\mu}}}$& $\pm$(0--0.39) &0 -- 2& 0.31--0.40 &-\\
&(IH)&$\sqrt{\frac{m_{e}}{m_{\mu}}}$& $\pm$(0.51--1) &0 -- 2& 0.31--0.40 &-\\
\hline
GR2&None& $\sqrt{2} \, s_{13}$& $\pm$(0.56--0.64) & $\pm$(0.51--0.72)&-&0.43-0.55  \\
\hline\hline
BM&& - & - & $\pm$(0.87--1)&- &0.44--0.56\\\hline\hline
&Data fit (NH) \cite{GonzalezGarcia:2012sz}&&0.9--2.03&0--2&0.29--0.31 (1$\sigma$)&0.408--0.414 (1$\sigma$)\\
& (IH) \cite{GonzalezGarcia:2012sz}&&0.9--2.03&0--2&0.29--0.31 (1$\sigma$)&0.57--0.61(1$\sigma$)\\
\hline
\end{tabular}
\label{results}
\end{table}

\subsection{Neutrino perturbations and $\s_{12}^{l}=0$}\label{8.1}
In this case the only correction from the charged lepton sector comes from $s_{23}^{l}$. From eqs(\ref{s13}) we have 
\bea
s_{13}&=&s_{13}^{\nu}=s_{13}^{\epsilon}\nonumber\\ 
\delta_{13}&=&\delta_{13}^{\nu}\nonumber\\
 \delta_{12}&=&\delta_{12}^{\nu}\nonumber\\
 \delta_{23}&=&\delta_{23}^{\nu}-s_{23}^{l}\sin\alpha/\sqrt{2}\nonumber\\
 s_{23}^{\nu2}&=& s_{23}^{2}+s_{23}^{l2}/2+\sqrt{2} s_{23}s_{23}^{l}\cos\alpha
 \label{lepcorr}
 \eea
 where $\alpha=\delta_{23}^{l}-\delta_{23}^{\nu}$ is unknown. 
  Thus
\be
 \delta\equiv \delta_{13} - \delta_{12} - \delta_{23}=\delta^{\epsilon}+s_{23}^{l}\sin\alpha/\sqrt{2}
 \ee
 and the correlated uncertainties in $\delta^{\epsilon}$ and $s_{23}^{\nu}$ are determined by $s_{23}^{l}$.

\subsubsection{$\nu_{a}\nu_{b}$ mixing}\label{secnuab}
$\bold{TBM}$

From eq(\ref{nuab}),
\be
\delta^{\nu}=\delta^{\epsilon}=\cos^{-1}(\frac{s_{23}^{\nu 2}-0.5-s_{13}^{2}}{\sqrt{2}s_{13}}).
\label{de}
\ee

    Using $s_{23}^{\nu}$ from eq(\ref{lepcorr}) we determine the allowed range of  the Dirac CP violating phase $\delta$ given the allowed range of $s_{23}$ from the fits of Table \ref{fits} and allowing for the uncertainty introduced by $\alpha$. The results are shown in the first two lines of Table \ref{results}.
  
 \vspace{0.3cm} 
\noindent $\bold{GR}$ 
 
 For the case of Golden ratio mixing eqs(\ref{nuab}) are replaced by eqs(\ref{nuabgr}) giving
 \be
\delta^{\nu}=\delta^{\epsilon}=\cos^{-1}(\frac{2s_{23}^{\nu 2}-1-\cot\Theta^2 s_{13}^{2}}{2s_{13}\cot\Theta}).
\label{degr}
\ee 
 Using $s_{23}^\nu$  from eq(\ref{lepcorr}) to include the uncertainty introduced by $s_{23}^l$, and taking $\tan \Theta = (1,2)/\phi$,  one obtains the prediction for $\delta$ in the (GR1, GR2) schemes shown in Table \ref{results}.

\subsubsection{$\nu_{a}\nu_{c}$ mixing}
$\bold{TBM}$

The difference compared to the case of $\nu_{a}\nu_{b}$ mixing is that eqs(\ref{nuab}) are replaced by eqs(\ref{nuac}). In practice the only change in the analysis is that eq(\ref{de}) is changed to 
\be
\delta^{\nu}=\delta^{\epsilon}=\cos^{-1}(-\frac{\sqrt{2}(s_{23}^{\nu 2}-0.5-s_{13}^{2}/4)}{s_{13}}).
\label{denew}
\ee 
The resulting prediction for $\delta$ is  given in Table \ref{results}.

 \vspace{0.3cm} 
\noindent $\bold{GR}$ 

In this case the only change is that from eq(\ref{nuacgr}) we have
\be
\delta^{\nu}=\delta^{\epsilon}=\cos^{-1}(-\frac{2s_{23}^{\nu 2}-1-s_{13}^{2}\tan^2\Theta}{2s_{13}\tan\Theta}).
\label{deacgr}
\ee 
\subsection{Neutrino perturbations and $\s_{12}^{l}=\sqrt{\frac{m_{e}}{m_{\mu}}}$}

A non-zero $\s_{12}^{l}$ affects the determination of $s_{13}^{\nu}$ and $s_{12}$. From eq(\ref{s13}) we have 
\bea
s_{13}^{\nu 2}&=&s_{13}^{2}+s_{12}^{l 2}/2+\sqrt{2}s_{13}s_{12}^{l}\cos\beta\nonumber\\
\delta_{13}^{\nu}&=&\delta_{13}+s_{12}^{l}\sin\beta/(\sqrt{2}s_{13})\nonumber\\
\eea
where $\beta=\delta_{23}^{\nu}+\delta_{12}^{l}-\delta_{13}$. Using these equations for given $s_{13}$ and $s_{12}^{l}$ we may determine $s_{13}^{\nu}$ and the uncertainty in the relation between $\delta_{13}^{\nu}$ and $\delta_{13}$ (and hence the error in the relation $\delta=\delta^{\epsilon}$) in terms of the unknown phase $\beta$ and thus estimate their (correlated) errors. 

Turning to $s_{12}$, eq(\ref{s12}) gives
\be
\sqrt{3}s_{12}e^{-i\delta_{12}}=e^{-i\delta_{12}^{\nu}}-s_{12}^{l}e^{-i\delta_{12}^{l}}.
\ee
We see that the prediction for $s_{12}$ depends on unknown phases so now there is a range of allowed values as shown in Table  \ref{fits}. Also, due to the unknown lepton phase, there is now an uncertainty in the relation between the neutrino contribution to the phase and the full phase, $\delta_{12}=\delta_{12}^\nu\pm s_{12}^{l}$, and this translates to an uncertainty in the determination of $\delta$ of about $2^{0}$. There is also an error coming from the unknown lepton phase, $\alpha$,  as discussed in Section \ref{8.1}.

Having allowed for these charged lepton contributions the analysis proceeds as detailed in Section \ref{8.1}. The predictions for the phases and angles are given in Table \ref{results}.

\subsection{No neutrino perturbations and $s_{12}^{l}= \sqrt{2} \, s_{13}$}
In this case $\theta_{13}$ is entirely given by the charged lepton sector. Given the size of $s_{12}^l$ it is necessary to keep the full dependence on it rather than use the approximate  eqs(\ref{s23}) giving (up to the correction from the uncertainty in $s_{23}^l$)

\bea
|s_{23}s_{12}-s_{13}c_{23}c_{12} e^{i\delta}|&=&\frac{s_{12}^{\nu}}{\sqrt{2}}\nonumber\\
s_{23}c_{13}&=&\frac{c_{12}^l}{\sqrt{2}}.
\label{exact}
\eea

The result for  $s_{23}$ is shown in Table \ref{results} where the error is dominated by the uncertainty in $s_{23}^l$ which we estimate as $m_{\mu}/m_{\tau}$.

\subsubsection{TBM} 
Currently the largest uncertainty in the above prediction for $\delta$ arises from the experimental uncertainties in $s_{12}$ and $s_{13}$, giving a range $\pm (0.54 - 0.61) \pi$ at $1 \sigma$ and $\pm (0.48 - 0.68) \pi$ at $3 \sigma$, as shown in Table \ref{results}.   Thus a pure TBM neutrino mass matrix together with a hierarchical charged lepton mass matrix leads to near maximal Dirac CP violation.   The experimental uncertainty in $\delta$ is currently dominated by the experimental uncertainty in $s_{12}^2$.  At $1 \sigma$ the fractional error in $s_{12}^2$ is about $\pm 0.06$ and this leads to an uncertainty of $\pm 0.03\pi$ in $\delta$.  A reduction in this uncertainty by a factor 2 would lead to a prediction of $\delta$ at the 5\% level. 
\subsubsection{GR mixing}
We find  the range $\pm (0.41-0.47)\pi$ at $1\sigma$ and $\pm (0.32-0.51)\pi$ at $3\sigma$ for GR1 and $\pm (0.56-0.64)\pi$ at $1\sigma$ and $\pm (0.51-0.72)\pi$ at $3\sigma$ for GR2. 

\subsection{BM mixing} From eq(\ref{BMM}) we may determine the mixing angles and Dirac phase. For $s_{23}$ the error is dominated by the leptonic contribution and, for $s_{23}^l=m_\mu/m_\tau$, this gives the range $s_{23}^2=0.44-0.56$.  BM mixing is inconsistent with the allowed range of $s_{12}^2$ at the $1\sigma$ level. At the $3\sigma$ level we find $\delta=(0.87-1)\pi$.

\section{More general mixing} \label{general}

Up to now we have only considered perturbations about pure TBM where a single mass-mixing operator, $\nu_a \nu_b, \nu_b \nu_c$ or $\nu_a \nu_c$, dominates.   Of course more general mixing is possible and we illustrate this with two examples.

\subsection{A model based on the flavor group $A_4$}
Pure TBM results in a model based on the flavor group $A_4$ having three flavon fields $\phi, \phi_S$ and $\phi_T$ and leading interactions \cite{Altarelli:2005yx}
\be
{\cal L} = \frac{1}{\Lambda} \Big( y_e (\phi_T l) e^c h_d +  y_\mu (\phi_T l)' \mu^c h_d +  y_\tau (\phi_T l)'' \tau^c h_d \Big) +  \frac{1}{\Lambda^2} \Big( x_a \phi (ll) + x_b (\phi_S ll) \Big) h_u h_u  \; + h.c.
\ee
The dominant perturbations to pure TBM result purely from the neutrino sector, and arise from three operators\cite{Altarelli:2005yx}
\be
{\cal L} =\frac{1}{\Lambda^3} \Big( x_c (\phi_T \phi_S)' (ll)'' +  x_d (\phi_T \phi_S)'' (ll)'  +  x_e \phi (\phi_Tll) \Big) h_u h_u  \; + h.c.
\ee
We find that the couplings $x_c$ and $x_d$ lead to only the $\nu_a \nu_c$ mass operator, while the $x_e$ couplings leads only to $\nu_b \nu_c$ mass mixing.  Hence in this model $\theta_{13}$ must arise from $x_c$ or $x_d$. If the deviation of $s_{12}$ from tri-maximal mixing is to be small as required by the data, the coefficient $x_e$ must be small. In this case the model reduces to a single mass-mixing operator. Dominance by more than one mass-mixing operator requires a correlation between the coefficients of the mass operators, as the next example demonstrates.
\subsection{A model that preserves $s_{23}^2 = 1/2$ and $s_{12}^2 = 1/3$}
Maintaining the tri-maximal solar angle and the maximal atmospheric angle requires 
\be
\frac{|U_{e2}|^{2}}{|U_{e1}^{2}|^{2}}=\frac{1}{2},\;\;\frac{|U_{\mu 3}|^{2}}{|U_{\tau3}^{2}|^{2}}=1
\ee
and occurs in the scheme  \cite{King:2009qt}
\be
{\cal L}_{M} = \frac{1}{2}m_3\left(\nu_{a}+\frac{\epsilon}{\sqrt{3}}(\nu_{b}+\sqrt{2}\nu_{c})\right)^{2}
+\frac{1}{2}m_2\left(\nu_{b}-\frac{\epsilon}{\sqrt{3}}\nu_{a}\right)^{2}
+\frac{1}{2}m_1\left(\nu_{c}-\sqrt{\frac{{2}}{{3}}}\epsilon\,\nu_{a}\right)^{2}  \; + h.c.
\label{correlation}
\ee
with simultaneous mixing  of $\nu_{b}$ and $\nu_{c}$ with $\nu_{a}$. This evades the correlation discussed above between the departure of $\s_{23}$ from TBM mixing and the value of $\s_{13}$\footnote{A drawback of such a scheme is that such a modification of the familon vevs  changes the leading order structure of the Dirac mass matrices of the charged leptons and quarks and thus spoils the phenomenologically successful mass predictions that follow from a (1,1) texture zero.   The bilinear mixing discussed in Section \ref{TBMperturbations} arises from higher order corrections to the Majorana mass matrix and thus does not spoil the (1,1) texture zero relations \cite{Varzielas:2012ss}.}. Although this requires a strong correlation between the mixing of these states it has been argued that this can happen quite naturally in an $A_{4}$ model through vacuum alignment of the familon vevs \cite{King:2009qt}. To see how this may come about note that the structure of eq(\ref{correlation}) follows from a modification of the structure discussed in Section \ref{SectionTBM}, namely
\be
{\cal L}_{M}= \frac{1}{2} m_3 (\nu_{i}\theta_{a}^{i})^{2}+ \frac{1}{2} m_2 (\nu_{i}\theta_{b}^{i})^{2} \; + h.c.
\ee
where now the triplets vevs have the form $\phi_a \propto (\epsilon,1,-1)$ and $\phi_{b}\propto (1,1,1)$.  The modification compared to the  tri-bi-maximal mixing scheme is the appearance of the entry proportional to $\epsilon$ in the vev of the familon field $\phi_{a}$. 

In the proposed alignment scheme there are additional triplet familon fields $\phi_{1},\;\phi_{3}$ and $\tilde{\phi}_{23}$. All the familon fields have vevs driven by radiative breaking through a potential of the form $V=m^{2}|\phi|^{2}$ where $m^{2}$ becomes negative at some high scale through radiative corrections. This term is $SU(3)$ symmetric so to determine the vacuum alignment one must look for potential terms splitting the degeneracy. An $A_{4}$ invariant that does this has the form $\sum_{i}|\phi^{\dagger i}\phi_{i}|^{2}$. If its coefficient is positive (negative) the preferred vev is $\propto (0,0,1)$ ($\propto(1,1,1)$).  It was argued that this mechanism readily leads to the vevs
\be
|<\phi_{123}|>\propto (1,1,1),\;\;<|\phi_{1}|>\propto (1,0,0),\;\;<|\phi_{3}|>\propto (0,0,1).
\ee

Finally the alignment terms 
\be
\tilde\lambda_{123}|\phi_{123}^{\dagger}\tilde\phi_{23}|^{2}+\tilde\lambda_{1}|\phi_{1}^{\dagger}\tilde\phi_{23}|^{2}+\lambda_{1}|\phi_{1}^{\dagger}\phi_{a}|^{2}+\lambda_{23}|\tilde\phi_{23}^{\dagger}\phi_{a}|^{2}
\label{alignment}
\ee
were added to try to obtain the desired alignment. The first two terms with positive coefficients force $<|\tilde\phi_{23}|>\propto(0,1,1)$. The last term forces $\tilde\phi_{23}$ and $\phi_{a}$ to be orthogonal but does not require the second and third terms of $\phi_{23}$ to be non-zero. This is determined by the third term - if $\lambda_{1}$ is positive the first term vanishes while if it is negative the second and third terms vanish.  Thus these terms do not drive the desired form of the vacuum alignment,  $\phi_a \propto (\epsilon,1,-1)$. To arrange for such alignment requires a modification of the alignment terms. The simplest possibility arises if $\phi_{1}$ and $\phi_{3}$ have the same quantum numbers under the symmetries beyond $A_{4}$ that are usually introduced to limit the allowed form of the Lagrangian. In this case the second and third terms  of eq(\ref{alignment}) can take the form\footnote{The most general structure is more complicated but these terms suffice to illustrate the point.} 
\be
\tilde\lambda_{1}|(\phi_{1}+x\phi_{3})^{\dagger}\tilde\phi_{23}|^{2}+\lambda_{1}|(\phi_{1}+y\phi_{3})^{\dagger}\phi_{a}|^{2}
\ee
where $x$ and $y$ are constants. Now it is straightforward to see that, for small $y$, the vev of $\phi_{a}$ has the desired form with $\epsilon=-y<\phi_{1}>/<\phi_{3}>$. However to avoid spoiling the alignment of $\tilde\phi_{23}$ it is necessary that $x$ be large and to keep the correcting within the oserved limits requires $y/x\le10^{-2}$. Given that one expects the coefficients $x$ and $y$ to be of $O(1)$ this looks an unnatural requirement. 

In summary, while it is possible to achieve the vacuum alignment necessary to achieve the form of eq(\ref{alignment}), this example shows that it requires a very complicated alignment mechanism and even with this requires some fine tuning of $O(1)$ coefficients. Thus, although it is possible to have more than a single mass-mixing operator, if the correction to tri-maximal mixing in the solar neutrino sector is to be naturally small, it is likely that a single mass-mixing operator should dominate and the structure analysed in the previous Sections apply.

\section {Summary and conclusions}\label{summary}
In this paper we have explored the possibility that the structure of the neutrino mixing matrix is given in zeroth order by the most promising family symmetries that have been suggested, namely those leading to TBM, GR or BM mixing and perturbed by mixing between the zeroth order mass eigenstates, $\nu_a,\;\nu_b$ and $\nu_c$. 

We discussed the underlying $Z_2\times Z_2$ family symmetries that lead to the unperturbed mixing and argued that, in order naturally to preserve the good prediction for $\theta_{12}$ in these schemes while generating an acceptably large value for $\theta_{13}$, it is necessary that the perturbations should leave a residual $Z_2$ factor unbroken. This  corresponds to the case that a single bilinear mixing term, $\nu_a\nu_b$ or $\nu_a\nu_c$ is dominant. We supported this contention by studying two examples which involved more general mixing and showed that, to avoid fine tuning,  they reduced to single bilinear dominance.

We also constructed a scheme that has only the atmospheric neutrino massive in the unperturbed case, while both the solar neutrino mass and $\theta_{13}$ are generated by first order mass perturbations. This has the merit of 
explaining the comparable magnitudes of the solar neutrino mass and the mass perturbation that generates $\theta_{13}$,
while preserving the phenomenology of the previous schemes.

To determine this phenomenology we developed the perturbative mixing analysis for TBM and for general Golden ratio schemes, and determined the resulting correlations  between the magnitude of $\theta_{13}$, the other mixing angles and the Dirac CP violating phase. In doing so, guided by the mixing in the quark sector,  we also allowed for a range of mixing in the charged lepton sector.  For comparison we also determined the correlations for the case that $\theta_{13}$ comes entirely from the charged lepton sector with the neutrino sector being given by pure TBM, GR or BM mixing. 

For $\nu_a\nu_b$ or $\nu_a\nu_c$ mixing perturbations, the correlations originate from 
\be
\cos \delta^\nu  \; \simeq \;  C \, \frac{1}{s^\nu_{13}} \left( s_{23}^{\nu 2} - \frac{1}{2} \right), \hspace{.5in} \mbox{with}\;\;  C = (\tan \theta_{12}^\nu, -\cot \theta_{12}^\nu) \;\;  \mbox{for} \;\; (ab, ac) \;\; \mbox{mixing}
\label{correlation}
\ee
to leading order in $s_{13}^\nu$.   Equivalently, these correlations also result from a residual $Z_2$ symmetry \cite{Ge:2011qn}.  To obtain a predicted range for the CP violating observable $\delta$ in terms of the measured mixing angles $\theta_{ij}$, this relation must be corrected with terms from the charged lepton sector involving $s_{23}^l$ and $s_{12}^l $.  The accuracy of this prediction is limited by both the experimental uncertainties of $\theta_{23,13}$ and by the unknown phases that enter the contributions from the charged lepton sector.  Using the present uncertainties on $\theta_{23,13}$, the results of our analysis are summarised in Table \ref{results}.  The differences between the predicted ranges of $\delta$ in the various schemes largely reflect the different values of $C$ in eq(\ref{correlation}).  The (TBM, GR1, GR2) schemes give $C=(0.71, 0.62, 1.24)$ for $ab$ mixing and $C=-(1.41,1.62,0.81)$ for $ac$ mixing.  Thus the ranges for TBM and GR1 are similar for both $ab$ and $ac$ mixing, while the GR2 prediction is more distinct.

Very significant improvements are expected in the measurements of $\theta_{12}$ and $\theta_{23}$ in the coming years.  In $ab$ and $ac$ mixing schemes, $\theta_{12}$ is unperturbed, so that a reduced uncertainty will tell us whether one of these schemes, with a small contribution from $\theta^l_{12}$, is allowed, as shown by the $s_{12}^2$ column of Table \ref{results}, and may serve to distinguish between TBM and GR schemes.  On the other hand, a reduction in the experimental uncertainty in $\theta_{23}$ will lead to a significant increase in precision in the predicted ranges of $\delta$, compared to the present predictions shown in Table \ref{results}; although a residual uncertainty from the unknown phases in the charged lepton contributions will remain.

There are two competing ideas for understanding the large neutrino mixing angles $\theta_{12}$ and $\theta_{23}$:  flavor symmetries and anarchy \cite{Hall:1999sn, Haba:2000be, deGouvea:2012ac}.   While the recent discovery of a relatively large value of $\theta_{13}$ has certainly increased the likelihood of anarchy, the competition is far from over.  Future precision neutrino experiments will greatly reduce the uncertainties on $\theta_{ij}$, determine whether the hierarchy is normal or inverted, and finally measure CP violation, providing it is not suppressed.  In this paper we have argued that such a program could yet uncover a very simple underlying structure of lepton flavor symmetry.

\section*{Acknowledgments}

One of us (GGR) would like to thank Pierre Ramond for useful discussions and the Leverhulme foundation for the award of an emeritus fellowship without which this research would not have been started. 
The work of LJH was supported in part by the Director, Office of Science, Office 
of High Energy and Nuclear Physics, of the US Department of Energy under 
Contract DE-AC02-05CH11231 and by the National Science Foundation under 
grant PHY-0855653.



\begin{thebibliography}{99}
\bibitem{Abe:2011sj}
{\bf T2K} Collaboration, K.~Abe {\em et.~al.}, {\it {Indication of Electron
  Neutrino Appearance from an Accelerator-Produced Off-Axis Muon Neutrino
  Beam}},  Phys. Rev. Lett. {\bf 107} (2011) 041801,
  arXiv:1106.2822

\bibitem{Adamson:2011qu}
{\bf MINOS} Collaboration, P.~Adamson {\em et.~al.}, {\it {Improved search for
  muon-neutrino to electron-neutrino oscillations in MINOS}},  Phys. Rev. Lett.
  {\bf 107} (2011) 181802, 
  arXiv:1108.0015

\bibitem{Abe:2011fz}
{\bf DOUBLE-CHOOZ} Collaboration, Y.~Abe {\em et.~al.}, {\it {Indication for
  the Disappearance of Reactor Electron Antineutrinos in the Double Chooz
  Experiment}}, 
  arXiv:1112.6353

\bibitem{An:2012eh}
{\bf DAYA-BAY} Collaboration, F.~P. An {\em et.~al.}, {\it {Observation of
  Electron-Antineutrino Disappearance at Daya Bay}},
 arXiv:1203.1669

\bibitem{Ahn:2012nd}
{\bf RENO} Collaboration, J.~K. Ahn {\em et.~al.}, {\it {Observation of Reactor
  Electron Antineutrino Disappearance in the Reno Experiment}},
  arXiv:1204.0626
  
\bibitem{GonzalezGarcia:2012sz}
  M.~C.~Gonzalez-Garcia, M.~Maltoni, J.~Salvado and T.~Schwetz,
  arXiv:1209.3023 [hep-ph].
  
  \bibitem{Fogli:2012ua}
G.~Fogli, E.~Lisi, A.~Marrone, D.~Montanino, A.~Palazzo, {\em et.~al.}, {\it
  {Global Analysis of Neutrino Masses, Mixings and Phases: Entering the Era of
  Leptonic CP Violation Searches}},
  arXiv:1205.5254

\bibitem{Tortola:2012te}
M.~Tortola, J.~Valle, and D.~Vanegas, {\it {Global Status of Neutrino
  Oscillation Parameters After Recent Reactor Measurements}},
 arXiv:1205.4018
 
\bibitem{Georgi:1979df}
  H.~Georgi and C.~Jarlskog,
  Phys.\ Lett.\ B {\bf 86} (1979) 297.
  
\bibitem{Gatto:1968zz}
  R.~Gatto, G.~Sartori and M.~Tonin,
  PRINT-68-2045.
  
\bibitem{Antusch:2011qg}
  S.~Antusch and V.~Maurer,
  Phys.\ Rev.\ D {\bf 84} (2011) 117301
  [arXiv:1107.3728 [hep-ph]].
  
\bibitem{Antusch:2012fb}
  S.~Antusch, C.~Gross, V.~Maurer and C.~Sluka,
  arXiv:1205.1051 [hep-ph].

\bibitem{Marzocca:2011dh}
  D.~Marzocca, S.~T.~Petcov, A.~Romanino and M.~Spinrath,
  JHEP {\bf 1111} (2011) 009
  [arXiv:1108.0614 [hep-ph]].
  
\bibitem{Hall:1993ni}
  L.~J.~Hall and A.~Rasin,
  Phys.\ Lett.\ B {\bf 315} (1993) 164
  [hep-ph/9303303].
\bibitem{Marzocca:2013cr} 
  D.~Marzocca, S.~T.~Petcov, A.~Romanino and M.~C.~Sevilla,
  arXiv:1302.0423 [hep-ph].
  
\bibitem{King:2013eh}
  S.~F.~King and C.~Luhn,
  arXiv:1301.1340 [hep-ph].
  
\bibitem{Altarelli:2010gt}
  G.~Altarelli and F.~Feruglio,
  Rev.\ Mod.\ Phys.\  {\bf 82} (2010) 2701
  [arXiv:1002.0211 [hep-ph]].

\bibitem{Ohlsson:2005js}
  T.~Ohlsson,
  Phys.\ Lett.\ B {\bf 622} (2005) 159
  [hep-ph/0506094].
  
\bibitem{Antusch:2005kw}
  S.~Antusch and S.~F.~King,
  Phys.\ Lett.\ B {\bf 631} (2005) 42
  [hep-ph/0508044].

\bibitem{King:2002nf}
  S.~F.~King,
  JHEP {\bf 0209} (2002) 011
  [hep-ph/0204360].
  
\bibitem{Ge:2011ih} 
  S.~-F.~Ge, D.~A.~Dicus and W.~W.~Repko,
  Phys.\ Lett.\ B {\bf 702}, 220 (2011)
  [arXiv:1104.0602 [hep-ph]].
  
\bibitem{Ge:2011qn} 
  S.~-F.~Ge, D.~A.~Dicus and W.~W.~Repko,
  Phys.\ Rev.\ Lett.\  {\bf 108}, 041801 (2012)
  [arXiv:1108.0964 [hep-ph]].
 

\bibitem{Harrison:2002er}
P.~F. Harrison, D.~H. Perkins, and W.~G. Scott, {\it {Tri-Bimaximal Mixing and
  the Neutrino Oscillation Data}},  Phys. Lett. {\bf B530} (2002) 167,
arXiv:hep-ph/0202074, hep-ph/0202074

\bibitem{Harrison:2002kp}
P.~F. Harrison and W.~G. Scott, {\it {Symmetries and Generalisations of
  Tri-Bimaximal Neutrino Mixing}},  Phys. Lett. {\bf B535} (2002) 163--169,
arXiv:hep-ph/0203209, hep-ph/0203209

\bibitem{Datta:2003qg}
 A.~Datta, F.~-S.~Ling and P.~Ramond,
 Nucl.\ Phys.\ B {\bf 671} (2003) 383
 [hep-ph/0306002].

\bibitem{Kajiyama:2007gx}
 Y.~Kajiyama, M.~Raidal and A.~Strumia,
 Phys.\ Rev.\ D {\bf 76} (2007) 117301
 [arXiv:0705.4559 [hep-ph]].

\bibitem{Rodejohann:2008ir}
  W.~Rodejohann,
  Phys.\ Lett.\ B {\bf 671} (2009) 267
  [arXiv:0810.5239 [hep-ph]].

\bibitem{Adulpravitchai:2009bg}
  A.~Adulpravitchai, A.~Blum and W.~Rodejohann,
  New J.\ Phys.\  {\bf 11} (2009) 063026
  [arXiv:0903.0531 [hep-ph]].
  
 \bibitem{Fukugita:1998vn}
  M.~Fukugita, M.~Tanimoto and T.~Yanagida,
  Phys.\ Rev.\ D {\bf 57} (1998) 4429
  [hep-ph/9709388].
  
\bibitem{Barger:1998ta}
  V.~D.~Barger, S.~Pakvasa, T.~J.~Weiler and K.~Whisnant,
  Phys.\ Lett.\ B {\bf 437} (1998) 107
  [hep-ph/9806387].
  
\bibitem{Davidson:1998bi}
  S.~Davidson and S.~F.~King,
  Phys.\ Lett.\ B {\bf 445} (1998) 191
  [hep-ph/9808296].
  
\bibitem{Feruglio:2011qq}
  F.~Feruglio and A.~Paris,
  JHEP {\bf 1103} (2011) 101
  [arXiv:1101.0393 [hep-ph]].
  
\bibitem{Blum:2007jz}
  A.~Blum, C.~Hagedorn and M.~Lindner,
  Phys.\ Rev.\ D {\bf 77} (2008) 076004
  [arXiv:0709.3450 [hep-ph]];
   A.~Blum, C.~Hagedorn and A.~Hohenegger,
  JHEP {\bf 0803} (2008) 070
  [arXiv:0710.5061 [hep-ph]].
  
\bibitem{Altarelli:2009gn}
  G.~Altarelli, F.~Feruglio and L.~Merlo,
  JHEP {\bf 0905} (2009) 020
  [arXiv:0903.1940 [hep-ph]].
  
\bibitem{King:2007pr}
  S.~F.~King,
  Phys.\ Lett.\ B {\bf 659} (2008) 244
  [arXiv:0710.0530 [hep-ph]].
  
 \bibitem{Albright:2008rp}
  C.~H.~Albright and W.~Rodejohann,
  Eur.\ Phys.\ J.\ C {\bf 62} (2009) 599
  [arXiv:0812.0436 [hep-ph]].

  \bibitem{leptonmixing}
S.~F.~King,   
  JHEP {\bf 0508} (2005) 105
  [hep-ph/0506297],

I.~Masina,   
  Phys.\ Lett.\ B {\bf 633} (2006) 134
  [hep-ph/0508031],

S.~Antusch, P.~Huber, S.~F.~King and T.~Schwetz,   
  JHEP {\bf 0704} (2007) 060 
  [hep-ph/0702286].

\bibitem{Xing:2001cx}
  Z.~-z.~Xing,
  Phys.\ Rev.\ D {\bf 64} (2001) 093013
  [hep-ph/0107005];
  C.~Giunti and M.~Tanimoto,
  Phys.\ Rev.\ D {\bf 66} (2002) 053013
  [hep-ph/0207096]; C.~Giunti and M.~Tanimoto,
  Phys.\ Rev.\ D {\bf 66} (2002) 113006
  [hep-ph/0209169];
  P.~H.~Frampton, S.~T.~Petcov and W.~Rodejohann,
  Nucl.\ Phys.\ B {\bf 687} (2004) 31
  [hep-ph/0401206].
  
  

\bibitem{Antusch:2009gu}
  S.~Antusch and M.~Spinrath,
  Phys.\ Rev.\ D {\bf 79} (2009) 095004
  [arXiv:0902.4644 [hep-ph]].
  
\bibitem{Altarelli:2005yx} 
  G.~Altarelli and F.~Feruglio,
  Nucl.\ Phys.\ B {\bf 741}, 215 (2006)
  [hep-ph/0512103].
  
\bibitem{King:2009qt}
  S.~F.~King,
  Phys.\ Lett.\ B {\bf 675} (2009) 347
  [arXiv:0903.3199 [hep-ph]].
  
\bibitem{Varzielas:2012ss}
  I.~de Medeiros Varzielas and G.~G.~Ross,
  arXiv:1203.6636 [hep-ph].

\bibitem{Hall:1999sn} 
  L.~J.~Hall, H.~Murayama and N.~Weiner,
  Phys.\ Rev.\ Lett.\  {\bf 84}, 2572 (2000)
  [hep-ph/9911341].
  
\bibitem{Haba:2000be} 
  N.~Haba and H.~Murayama,
  Phys.\ Rev.\ D {\bf 63}, 053010 (2001)
  [hep-ph/0009174].
  
\bibitem{deGouvea:2012ac} 
  A.~de Gouvea and H.~Murayama,
  arXiv:1204.1249 [hep-ph].

\end{thebibliography}
\end{document}